\documentclass{sig-alternate-2013}
\usepackage[top=1in, left=1in, right=1in, bottom=1.75in]{geometry}
\usepackage{epsfig}
\usepackage{citesort}
\usepackage{amsmath}
\usepackage{amssymb}
\usepackage{color}
\usepackage{url}
\usepackage{array}
\usepackage{tabu}
\usepackage[draft,colorlinks=true,urlcolor=black]{hyperref}

\usepackage{graphicx}
\usepackage{enumerate}
\usepackage{csquotes}
\usepackage{todonotes}
\usepackage{relsize}
\usepackage{caption}
\usepackage[labelfont=bf]{caption}
\usepackage{subcaption}
\usepackage{booktabs}
\usepackage{float}
\usepackage{cite}
\usepackage{flushend}
\usepackage{multirow}
\usepackage{hhline}
\usetikzlibrary{automata,topaths}
\usepackage{mathtools}

\usepackage{pgfplots}

\def\addlegendimage{\csname pgfplots@addlegendimage\endcsname}

\newcommand{\comm}[1] {
}
\makeatletter
\def\@copyrightspace{\relax}
\makeatother
\begin{document}
\title
{\large
    \textbf{5.4} \hspace{-0.5mm} \textbf{SMALL POLYGON COMPRESSION FOR INTEGER COORDINATES}
}
\setlength{\columnsep}{0.25in}
\def\Inf{\operatornamewithlimits{inf\vphantom{p}}}

\author{%
{\small\begin{minipage}{\linewidth}\begin{center}
\begin{tabular}{c}
    Abhinav Jauhri, Martin Griss \& Hakan Erdogmus\thanks{\textit{Corresponding author address:} Abhinav Jauhri, Carnegie Mellon University, ECE Department, Moffett Field, CA 94035; email: ajauhri@cmu.edu} \\
Carnegie Mellon University, Moffett Field, California \\
\end{tabular}
\end{center}\end{minipage}}
}
  
\maketitle
\pagestyle{plain}
\pagenumbering{arabic}
\begin{abstract}
We describe several polygon compression techniques to enable efficient transmission of polygons representing geographical targets. The main application is to embed compressed polygons to emergency alert messages that have strict length restrictions, as in the case of Wireless Emergency Alert messages. We are able to  compress polygons to between 9.7\% and 23.6\% of original length, depending on characteristics of the specific polygons, reducing original polygon lengths from  43-331 characters to 8-55 characters. The best techniques apply several heuristics to perform initial compression, and then other algorithmic techniques, including higher base encoding. Further, these methods are respectful of computation and storage constraints typical of cell phones.  Two of the best techniques include a \enquote{bignum} quadratic combination of integer coordinates and a variable length encoding, which takes advantage of a strongly skewed polygon coordinate distribution. Both techniques applied to one of two \enquote{delta} representations of polygons are on average able to reduce the size of polygons by some 80\%. A repeated substring dictionary can provide further compression, and a merger of these techniques into a \enquote{polyalgorithm} can also provide additional improvements.
\end{abstract}

\section{Introduction}\label{sec:intro}
Geo-targeting is widely used on the Internet to better target users with advertisements, multimedia content, and essentially to improve the user experience. Such information helps in marketing brands and increasing user engagement. Scenarios also exist where geo-targeting at a given time becomes imperative for specific sets of people for information exchange, thereby contributing to problems of network congestion and effectiveness. Emergency scenarios are a quintessential example where people in the affected area need to be informed and guided throughout the duration of an emergency. To address this, Wireless Emergency Alerts (WEA) is a nation-wide system for broadcasting short messages\footnote{\url{https://www.fema.gov/wireless-emergency-alerts}} (currently 90 characters, similar to SMS messages) to all phones in a designated geographic area via activation of appropriate cell towers. The area is typically identified by a polygon, though currently many operators use rather coarse-grained targeting (such as to a whole county).

Our research group has developed and evaluated improved geo-targeting technology. For testing purposes, we ran trials of the new system, called by us WEA+. We are using SMS (and WiFi)  to simulate cell broadcast, and also have on campus an experimental cell system, CROSSMobile\cite{iannucci2014crossmobile}, that supports true cell broadcast on an unused GSM frequency to cell phones with a special SIM card.

In order to do this, we included some compressed polygon representation as part of the short message text, expected to be feasible in both current 90 WEA character messages and even more effective in future implementations of WEA that allow longer messages, or use multiple messages. We have available to us a corpus of 11,370 WEA messages sent out by the National Weather Service (NWS)\cite{noaadataset}\footnote{These messages were sent out by the NWS in 2012, 2013, and through December, 2014}. The polygons in the NWS corpus range from 4-24 points, with a size ranging from 43-331 characters.  Since WEA messages are broadcast to thousands of people,  and it is believed that adding a polygon to more precisely define the target area is critical, it is essential to be able to compress typical WEA polygons to fit within the current or anticipated future WEA message length, leaving room for meaning text as well.  In this paper, we explore several ways of substantially compressing such polygons using heuristics and standard algorithms. Our techniques  provide better compression for almost all polygons in the corpus than standard algorithms. 

\begin{figure*}[ht!]
    \centering
    \includegraphics[width=150mm]{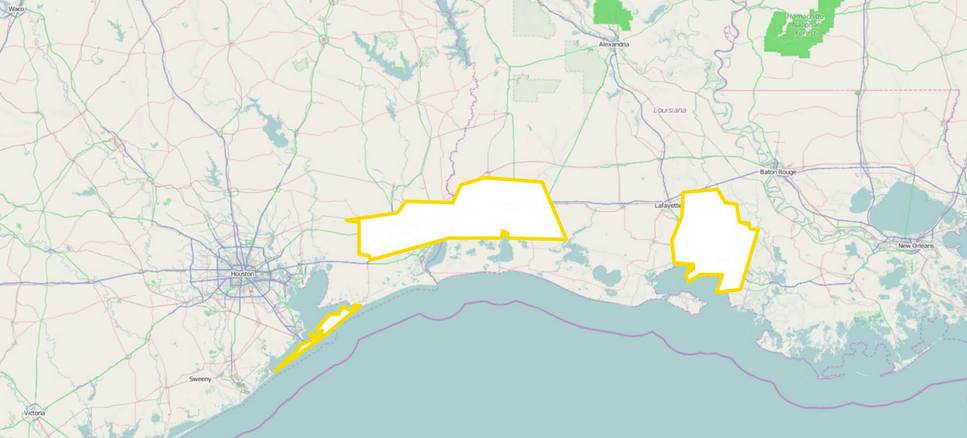}
    \caption{A map showing 3 polygons (yellow border). The NWS dataset has both convex and non-convex polygons}
    \label{wea_map}
\end{figure*}
\vspace{4mm}

The compression problem we are tackling here is quite different from that described in most other published research on polygon compression \cite{alliez2005recent,gandoin2002progressive}. They typically are dealing with a large number of inter-connected polygons in a 2D or 3D representation of a surface or solid, and thus are compressing a large number of polygons at the same time. Many of these polygons share common points and edges, which can be exploited in the compression; in our case, we have a single, relatively small polygon to compress, and so can not amortize items such as a dictionary of common points.

\begin{figure*}
\captionsetup{justification=centering}
\centering
\begin{subfigure}{.45\textwidth}
  \centering
  \includegraphics[width=\linewidth]{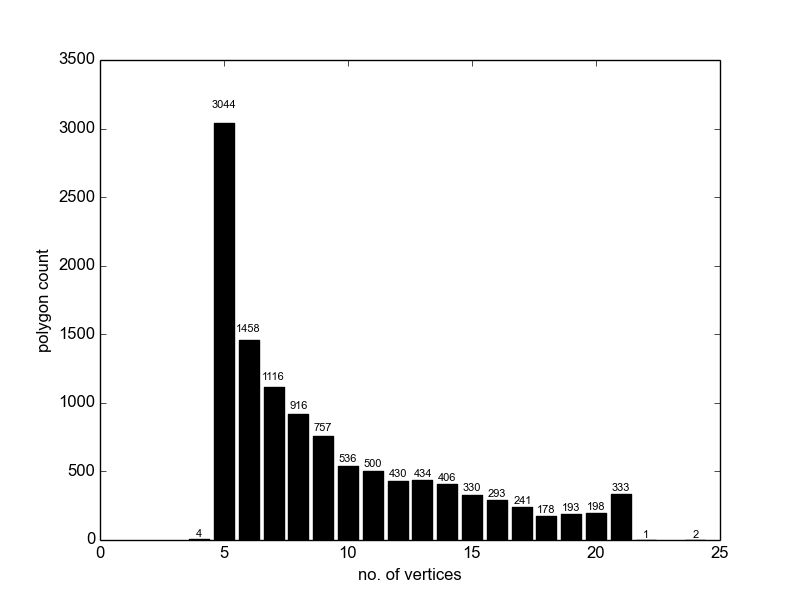}
  \caption{Distribution of polygon vertices}
  \label{polygon_vertices}
\end{subfigure}%
\begin{subfigure}{.45\textwidth}
  \centering
    \includegraphics[width=\linewidth]{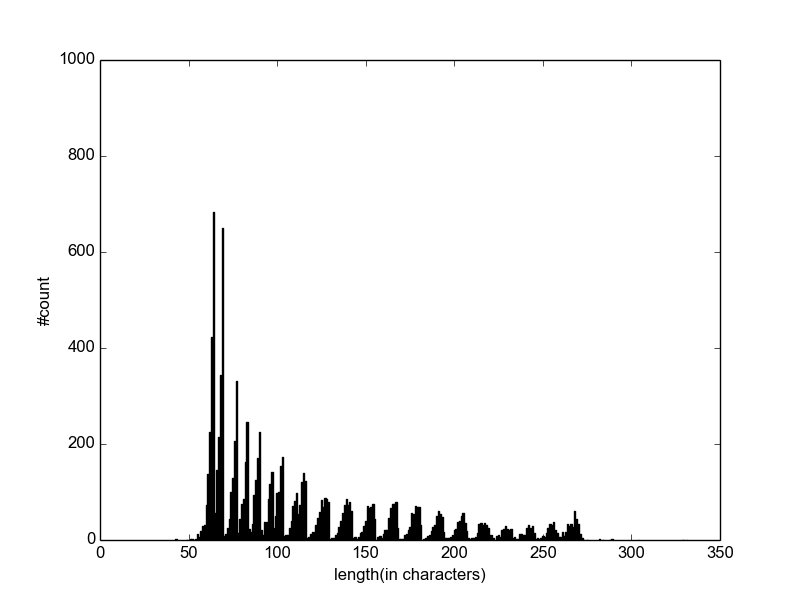}
    \caption{Distribution of original polygon lengths}
    \label{polygon_lengths}
\end{subfigure}
\caption{Distribution of NWS data}
\end{figure*}

In designing our techniques, we have looked at the typical distribution of coordinate values, polygon sizes and character string length of original uncompressed polygons, shown in Figures \ref{polygon_vertices} and \ref{polygon_lengths}. Many of these characteristics have motivated our transformations of the original polygon.

In the NWS corpus the observed range of GPS coordinates, covering a significant portion of the USA, is: 
        \[(X,Y)= (17.67,-159.32) ~\text{to}~ (48.84, -64.56)\]
        
We use three types of transformations on the numeric strings representing a polygon. The first group exploits heuristics, redundancies and patterns in the GPS coordinates, substantially simplifying and reducing the number of numeric characters in most polygon coordinates. The next set of transformations encodes the now simplified coordinates in a higher base, $B$,  such as all alpha-numeric characters, and then applies arithmetic and character string operations to further compress each polygon. The final set of transformations uses the statistical nature of the entire set of polygons to further compress some polygons.

We evaluated our corpus of polygons with combinations of the following compression techniques:
\begin{enumerate}
\item Purely heuristic, using deltas and fixed or variable length fields
\item Encoding in a higher base
\item Using a repeated substring dictionary to replace most frequent occurring substrings
\item Arithmetic operations to combine values
\item Arithmetic encoding, an entropy based compression technique
\item Other standard algorithms - LZW \cite{matt}, \textit{7zip}, \textit{gzip}, \textit{Huffman} \cite{huffman1952method} and \textit{Golomb} \cite{golomb}
\end{enumerate}

A key constraint is to compress the polygon into a string of characters that are acceptable via SMS or cell broadcast and specifically to the gateways we use to send an SMS via email for our pilot trials. To compress the set of decimals (or bits) representing GPS coordinates, we  use a base $B$ representation, avoiding characters that are questionable. This is discussed further in the practical considerations section. Base $B=62$ is a convenient choice since it uses only alphanumeric characters [0-9a-zA-Z]. Using a higher base such as $B=70$ or $B=90$ uses more characters and will improve the overall compression, and changes the trade-off between techniques. Briefly, our paper explores numerous techniques of compression on different transformations, or manipulations to the original polygon. 

%del - o178 csv75 big27 var30
%diff- o178 csv64 big25 var28

\section{Heuristic Approaches} \label{sec:heuristics}
We have combined several heuristics, motivated by analysis and discovered experimentally to work well. The following describes our current techniques. We start with an original $N$ point polygon, given as an ordered finite sequence in $\mathbb R^2$:\\
\[O=[X_1, Y_1, \ldots , X_N,Y_N]\] 
where $X_i, Y_i$ are GPS coordinates in decimal degrees.  

% Hakan: O looks like a set, but it isn't. Is order not important here? It should be because a polygon is not an arbitrary subset of Reals. If it is important: then I'd omit braces everywhere you use them.

In the NWS corpus, $N \in [4,24]$, even though the NWS standard allows polygons of up to 100 points. The original uncompressed polygon length of 43 to 331 characters includes $2N - 1$ separating commas, $2N$ periods and $N$ minus signs. Since $X_i$ is typically $dd.dd$ and $Y_i$ is typically $-dd.dd$ or $-ddd.dd$, the total length of the original polygon string $O$ is:

{\small
\begin{equation}\label{eq:orig_len}
\begin{split}
len(O) &<= \underbrace{4N}_{X_i} + \underbrace{5N}_{Y_i} + \underbrace{2N - 1}_\text{commas} + \underbrace{2N}_\text{periods} + \underbrace{N}_\text{minuses} \\
&<= 14N - 1
\end{split}
\end{equation}}

There are several steps we take to successively compress the polygon. Initially we perform three simplifications and transformations to the original set of coordinates $O$. The first transformation converts all coordinates to positive integers to yield, $O^\prime$. The second finds the minimum x-coordinate and y-coordinate and takes the difference with every  vertex ($T^{\Delta min}$). The third transformation considers the difference of consecutive coordinates ($T^{\Delta}$). \\

\begin{figure*}
\captionsetup{justification=centering}
\centering
\begin{subfigure}{.4\textwidth}
  \centering
  \includegraphics[width=0.85\linewidth]{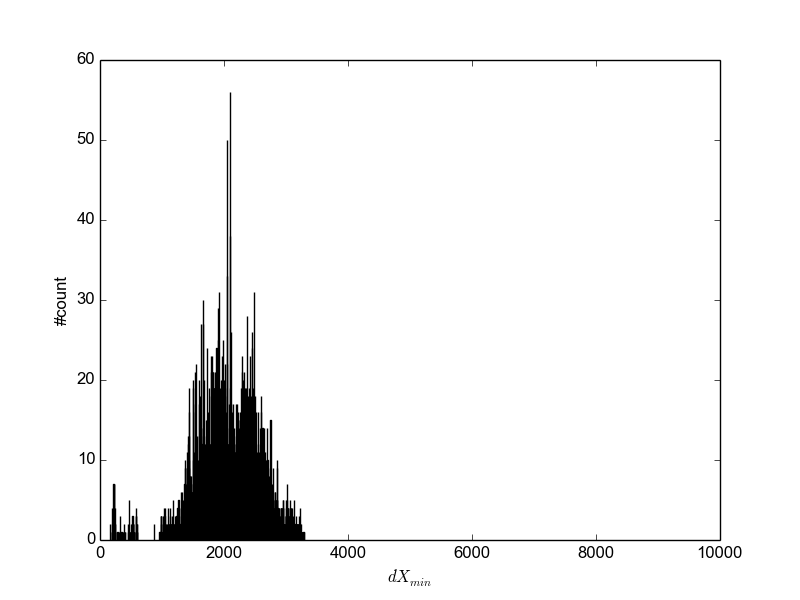}
    \caption{$dX_{min} \in [167, 3284]$}
    \label{fig:dxmin_distribution}
\end{subfigure}%
\begin{subfigure}{.4\textwidth}
  \centering
    \includegraphics[width=0.85\linewidth]{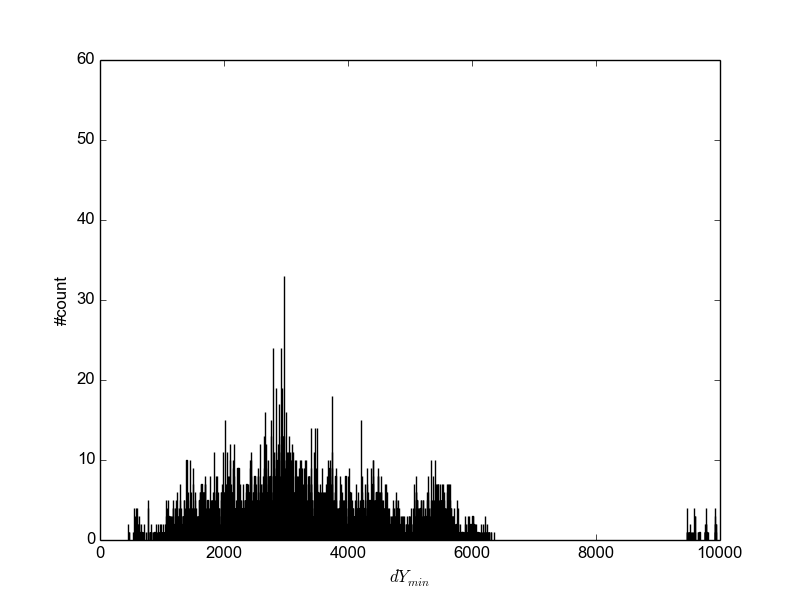}
    \caption{$dY_{min} \in [456, 9932]$}
    \label{fig:dymin_distribution}
\end{subfigure} \\
\begin{subfigure}{.4\textwidth}
  \centering
  \includegraphics[width=0.85\linewidth]{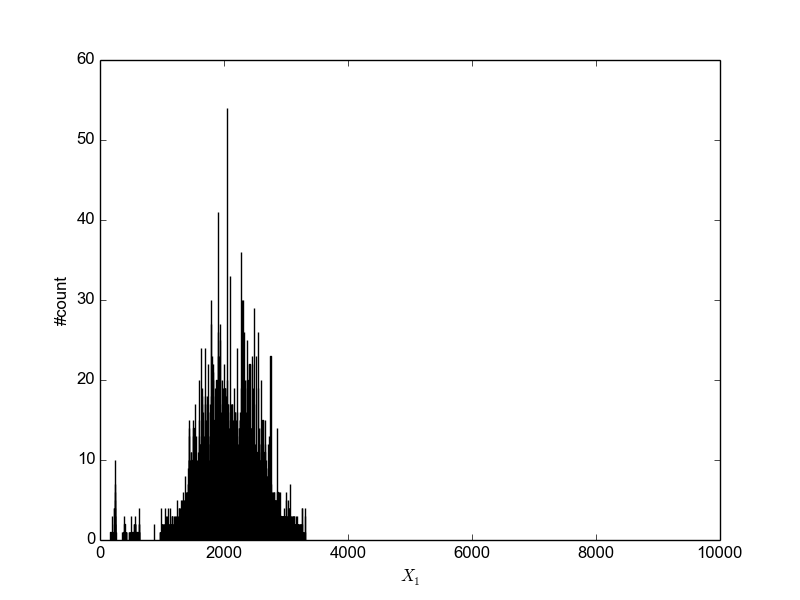}
    \caption{$\delta X_1 \in [169, 3301]$}
    \label{fig:x1_distribution}
\end{subfigure}%
\begin{subfigure}{.4\textwidth}
  \centering
    \includegraphics[width=0.85\linewidth]{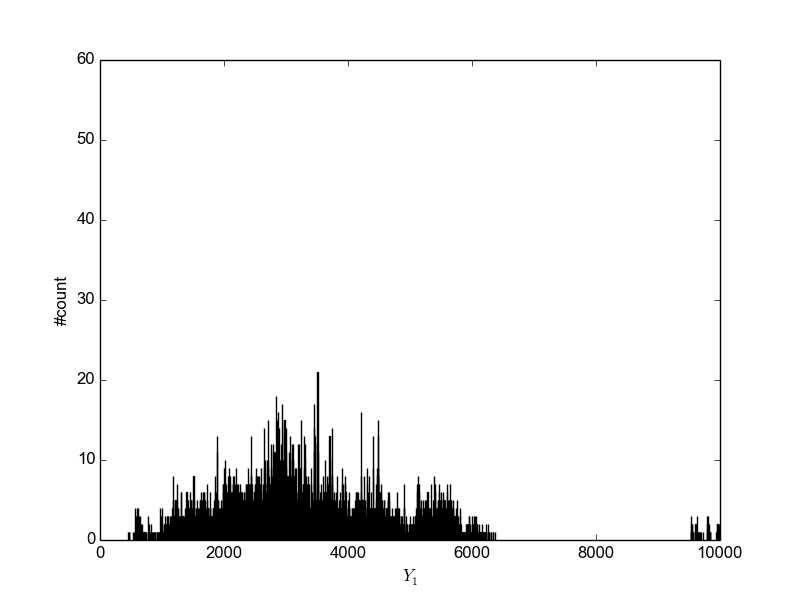}
    \caption{$\delta Y_1 \in [458, 9988]$}
    \label{fig:y1_distribution}
\end{subfigure}
    \caption{Ranges for large values in both transformations. $T^{\Delta min}$ has ($dX_{min}, dY_{min})$, and $T^\Delta$ has ($\delta X_1$, $\delta Y_1$)}
    \label{fig:large_distribution}
\end{figure*}

\begin{enumerate}[Step 1:]
\item Starting with polygon $O$, round all numbers to 2 (or 3) decimals precision,  convert to integers to drop the decimal point, and switch sign of $Y_i$ in USA, so both $X_i$ and $Y_i$ are positive integers, to produce $O^\prime$ \footnote{Following Mike Gerber of the NWS \cite{mike}}:
\[X_i = int(100 \ast X_i); Y_i = - int(100 \ast Y_i)\]
Outputting these with $2N-1$ separating commas gives a length of at most $11N-1$ characters as opposed to the original $14N-1$ (refer Eq. \ref{eq:orig_len}).
\item Compute $X_{min} = \Inf_i X_i,~ Y_{min} = \Inf_i Y_i$.
\item Compute deltas for all coordinates:
    \[dX_i=X_i-X_{min}\]
    \[dY_i=Y_i-Y_{min}\]
where $dX_i$ and $dY_i$ are non-negative integers.
\item Compute deltas for  $X_{min}$ and $Y_{min}$ from a chosen ``origin'',  origin $(X_0, Y_0)$:
\[dX_{min} = X_{min} - X_0\]
\[dY_{min} = Y_{min} - Y_0\]
We found $(1600, 6000)$ most effective (see Figures \ref{fig:dxmin_distribution} and \ref{fig:dymin_distribution}).
\item Since these are closed polygons, drop the last point $(X_N,Y_N)$ which is a duplicate of the first point, producing a shorter set of coordinates:
    \[T^{\Delta min}=[dX_{min}, dY_{min}, dX_1, dY_1,\ldots,\] 
    \vspace{-5mm} \[\ldots, dX_{N-1}, dY_{N-1}]\]
\end{enumerate}

\begin{figure*}[t]
\captionsetup{justification=centering}
\centering
\begin{subfigure}{.4\textwidth}
  \centering
  \includegraphics[width=0.85\linewidth]{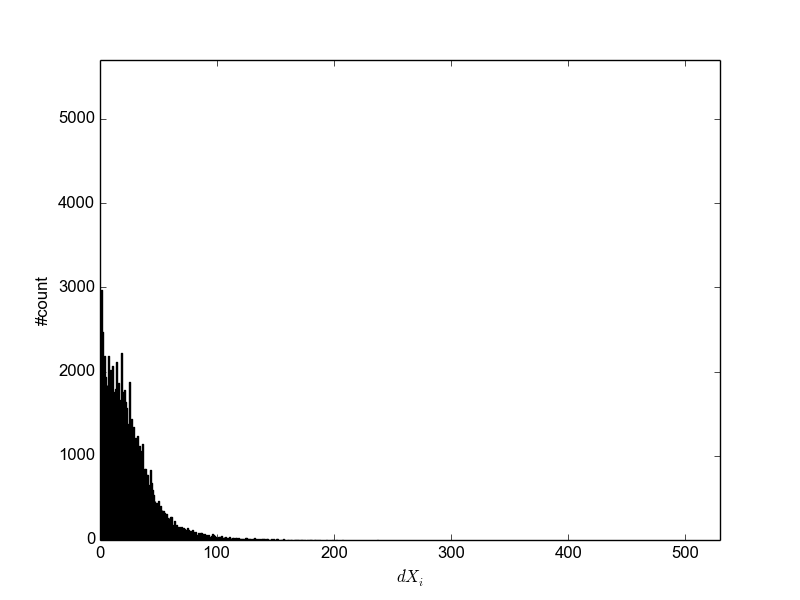}
  \caption{$skewness=1.63$; $kurtosis=1.42$ \\ $dX_i \in [1,327]$}
  \label{fig:delta_distribution_x}
\end{subfigure}%
\begin{subfigure}{.4\textwidth}
  \centering
    \includegraphics[width=0.85\linewidth]{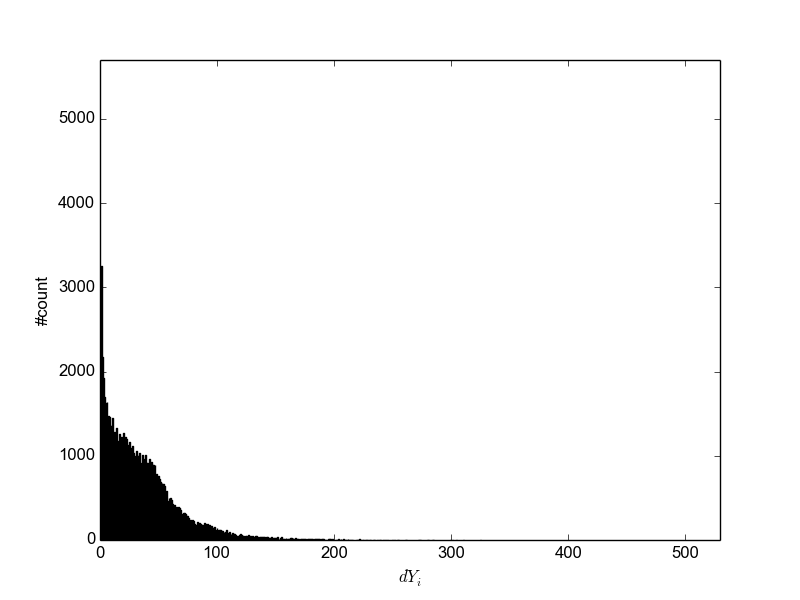}
    \caption{$skewness=2.08$; $kurtosis=5.34$ \\ $dY_i \in [1,325]$}
    \label{fig:delta_distribution_y}
\end{subfigure}\\
\begin{subfigure}{.4\textwidth}
  \centering
  \includegraphics[width=0.85\linewidth]{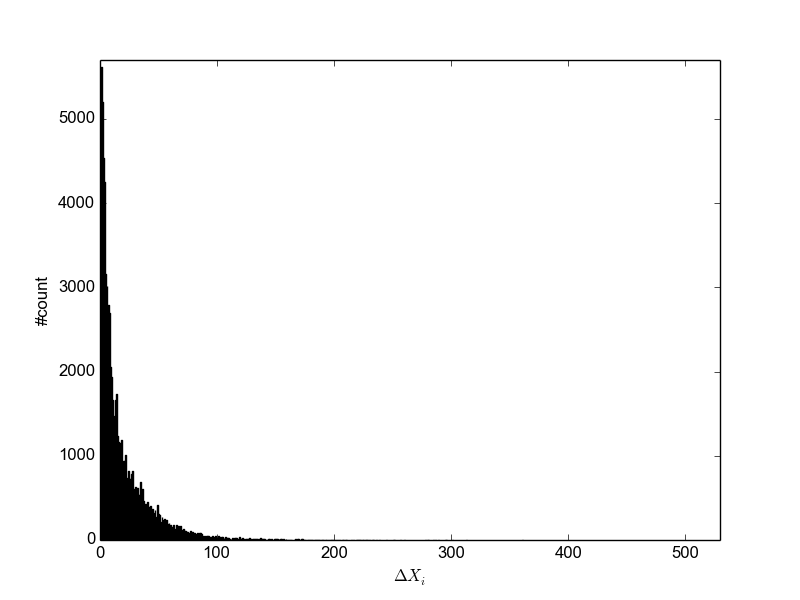}
  \caption{$skewness=4.34$; $kurtosis=21.17$ \\ $\Delta X_i \in [1,361]$}
  \label{fig:diff_distribution_x}
\end{subfigure}%
\begin{subfigure}{.4\textwidth}
  \centering
    \includegraphics[width=0.85\linewidth]{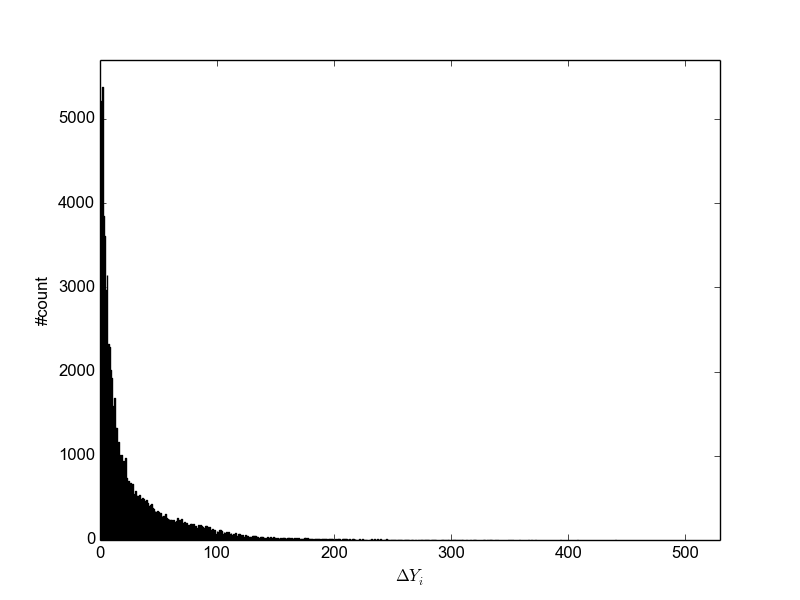}
    \caption{$skewness=5.2$; $kurtosis=31.56$ \\ $\Delta Y_i \in [1,524]$}
    \label{fig:diff_distribution_y}
\end{subfigure}
\caption{Positive skewness in the deltas $dX_i$(\ref{fig:delta_distribution_x}) and $dY_i$(\ref{fig:delta_distribution_y}). Heavy tails and peakedness define positive kurtosis \cite{decarlo1997meaning}. $kurtosis=85.25$ and $139.76$ for $dX_i$ and $dY_i$ respectively if 14354 cases where $dX_i = 0$ and 13838 cases where $dY_i = 0$ are included. Kurtosis not very high with inclusion of $\Delta X_i=0$ and $\Delta Y_i=0$ since their distributions have uniform fall off.}
\label{fig:delta_diff_distribution}
\end{figure*}

Steps for the second transformation $T^\Delta$ are very similar:

\begin{enumerate}[Step 1:]
\item Round all numbers to 2 (or 3) decimals precision,  convert to integers to drop the decimal point, and switch signs for $Y_i$:
\[X_i = int(100 \ast X_i); Y_i = -int(100 \ast Y_i)\]
\item Compute deltas for all coordinates:
    \[\Delta X_{i+1}=X_{i+1} - X_{i}\]
    \[\Delta Y_{i+1}=Y_{i+1} - Y_{i}\]
\item Compute deltas for  $X_{1}$ and $Y_{1}$ from a the chosen ``origin'', $(X_0, Y_0)$ :
\[\delta X_{1} = X_{1} - X_0\]
\[\delta Y_{1} = Y_{1} - Y_0\]
Here again we used $(1600, 6000)$ as ``origin''.
\item Many of the  $\Delta$s are negative integers which causes problems for the compression techniques discussed below. Therefore, every $\Delta X_i$ or $\Delta Y_i$ element $e$ will be converted as follows:
\[
    e=
    \begin{cases}
      2e, & \text{if}\ e>=0 \\
      -2e-1, & \text{if}\ e<0 \\
    \end{cases}
  \]
\item Drop the last point $(X_N,Y_N)$ which is a duplicate of the first point, producing a shorter set of coordinates:
\[T^{\Delta}= [\delta X_1,\delta Y_1,\Delta X_2,\Delta Y_2, \ldots , \Delta X_{N-1}, \Delta Y_{N-1}]\].
\end{enumerate}

Note that in addition to dropping the last point, we also save an additional point, since we start from $\delta X_1$ and $\delta Y_1$ rather than the additional $X_{min}$, $Y_{min}$ in the $T^{\Delta min}$ transformation. This form is particularly interesting, since the distribution of its $\Delta X_i$ has a skewed but less peaked shape to that of the $dX_i$ in $T^{\Delta min}$, but with a longer tail (See Figure \ref{fig:delta_diff_distribution}).

These heuristics already produce a substantial compression  because of the limited ranges and skewed distribution of the delta polygon coordinates ($dX_i$, $dY_i$), and ($\Delta X_i$, $\Delta Y_i$)  and of the starting points ($dX_{min}, dY_{min}$), and ($\delta X_1$, $\delta Y_1$), shown in Figures \ref{fig:delta_diff_distribution}, \ref{fig:large_distribution}. While the effectiveness of the various compression techniques work well because of these range and skew characteristics,  many of the techniques are not strongly dependent on the specifics, and thus many would work well even for a somewhat different set of polygons.

It is important to note that in both forms of the delta transformations $T$ we have two different sets of integers to compress, with distinctly different ranges and distributions: the single starting point pair of $dX_{min}$ and $dY_{min}$ for $T^{\Delta min}$ (or $\delta X_{1}$ and $\delta Y_{1}$ for $T^\Delta$) and the $N-1$ pairs $dX_{i}$, $dY_i$ for $T^{\Delta min}$ (or $N-2$ pairs $\Delta X_{i}$ and $\Delta Y_{i}$ for $T^\Delta$). We will thus treat them separately to get the best results.

As we shall see, for the NWS corpus two of the techniques are best, but we describe several others and the sub-transformations since some of these might perform better for other sets of polygons.

As a very first step to compress polygon, $T^{\Delta min}$ or $T^\Delta$ could be directly encoded as a comma-delimited string:
{\small
\begin{multline}
        T^{\Delta min}_c=dX_{min} \bullet,\bullet dY_{min} \bullet,\bullet dX_1 \bullet,\\\bullet dY_1 \bullet,\bullet \ldots \bullet dX_{N-1} \bullet,\bullet dY_{N-1}
    \end{multline}
    \begin{multline}
T^{\Delta}_c=\delta X_{1} \bullet,\bullet \delta Y_{1} \bullet,\bullet \Delta X_2 \bullet, \\ \bullet \Delta Y_2 \bullet,\bullet \ldots \bullet \Delta X_{N-2}\bullet,\bullet \Delta Y_{N-2}
\end{multline}}

\begin{figure*}
\captionsetup{justification=centering}
\centering
\begin{subfigure}{.4\textwidth}
  \centering
  \includegraphics[width=0.85\linewidth]{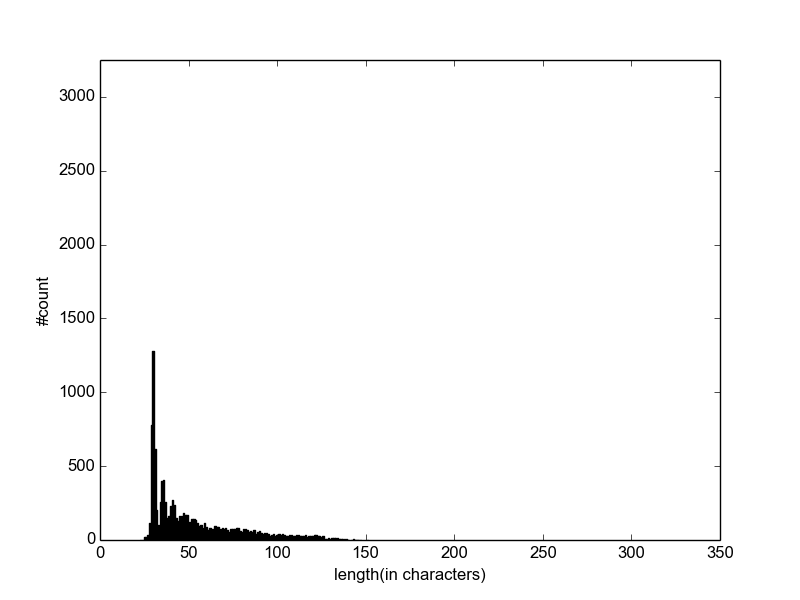}
    \caption{Distribution of $T^{\Delta min}_c$ lengths}
    \label{fig:polygon_del_c_length}
\end{subfigure}%
\begin{subfigure}{.4\textwidth}
  \centering
    \includegraphics[width=0.85\linewidth]{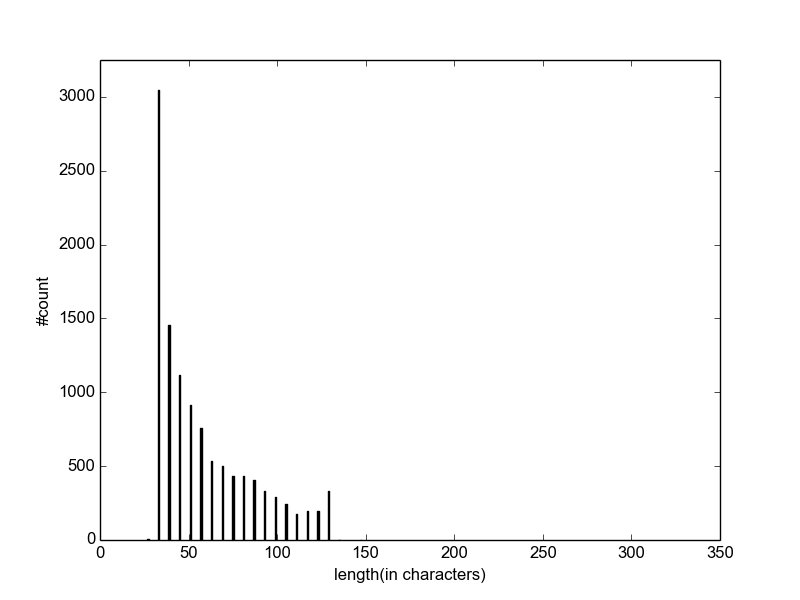}
    \caption{Distribution of $T^{\Delta min}_f$ lengths}
    \label{fig:polygon_del_f_length}
\end{subfigure}\\
\begin{subfigure}{.4\textwidth}
  \centering
  \includegraphics[width=0.85\linewidth]{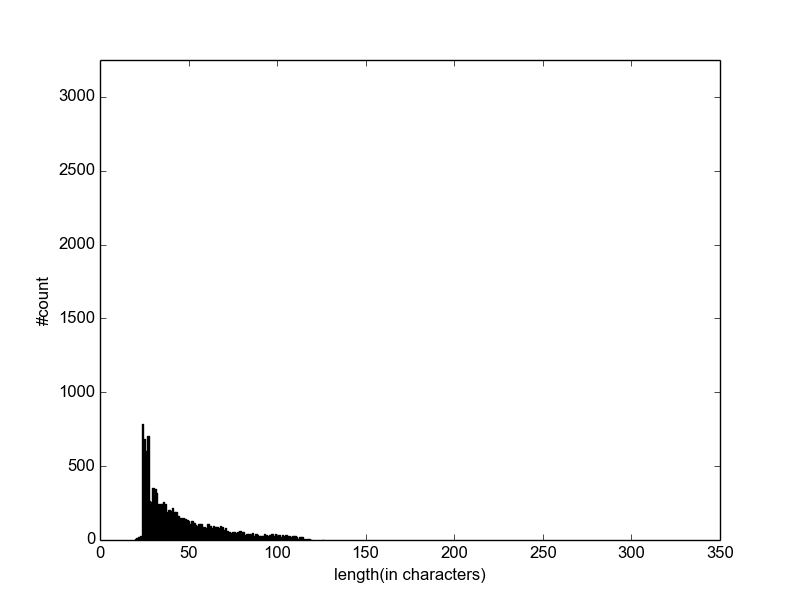}
    \caption{Distribution of $T^{\Delta}_c$ lengths}
    \label{fig:polygon_diff_c_length}
\end{subfigure}%
\begin{subfigure}{.4\textwidth}
  \centering
    \includegraphics[width=0.85\linewidth]{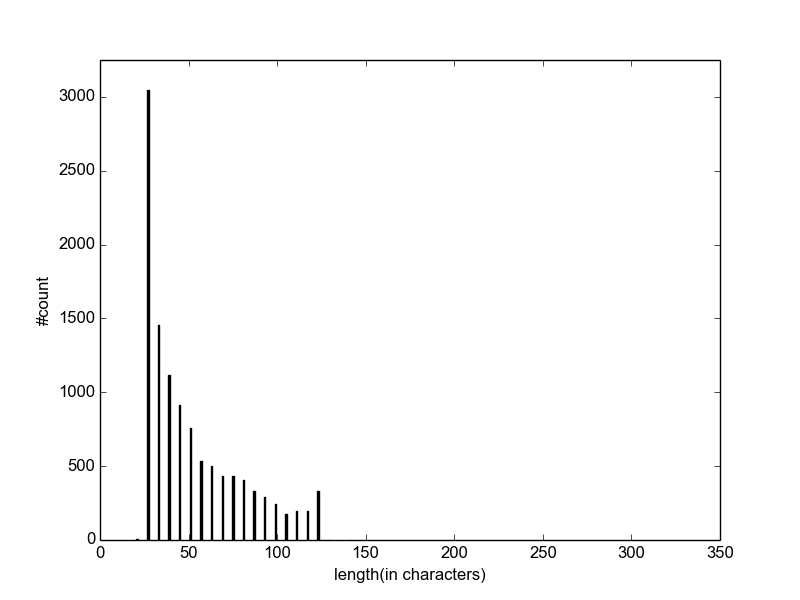}
    \caption{Distribution of $T^{\Delta}_f$ lengths}
    \label{fig:polygon_diff_f_length}
\end{subfigure}
    \caption{Distribution of base 70 compressed lengths for both transformations}
    \label{fig:polygon_comparisons}
\end{figure*}

Symbol $\bullet$ is used to denote string concatenation. Figure \ref{fig:polygon_comparisons} shows the distribution of lengths using this transformation. Since all deltas, $dX_i$, $dY_i$ are less than 350, (and $\Delta X_i$, $\Delta Y_i$ are less than 550) each of these deltas can be encoded in at most three decimal digits, $ddd$, while $dX_{min}$ or $\delta X_1$ will take at most four digits, $dddd$, and $dY_{min}$ or $\delta Y_1$ at most five digits, $ddddd$. The lower bound on the length of $T^{\Delta min}_c$ is thus:
{\small
\begin{equation}
\begin{split}
len(T^{\Delta min}_c) &>= \underbrace{1}_{dX_{min}} + \underbrace{1}_{dY_{min}} + \underbrace{(N - 1)}_{dX_i} + \underbrace{(N - 1)}_{dY_i} + \underbrace{2N - 1}_\text{commas} \\
&>= 4N - 1
\end{split}
\end{equation}}
and $4N-5$ for $T_c^\Delta$, 
though it is very rare in the NWS corpus for $dX_{min}$ and $dX_{min}$ to be encodable in a single digit.

The upper bound on the length of $T^{\Delta min}_c$ is:
{\small
\begin{equation}
\begin{split}
len(T^{\Delta min}_c) &<= \underbrace{4}_{dX_{min}} + \underbrace{5}_{dY_{min}} + \underbrace{3(N - 1)}_{dX_i} + \underbrace{3(N - 1)}_{dY_i}\\
& \hspace{10mm} + \underbrace{2N - 1}_\text{commas} \\
&<= 8N + 2
\end{split}
\end{equation}
}
and $8N-6$ for $T_c^\Delta$.

We can eliminate all commas by using fixed three digit fields for each delta, padded with zero, four digits for $dX_{min}$ and $\delta X_1$ and five digits for $dY_{min}$ and $\delta Y_1$ to get a significant improvement (Figure  \ref{fig:polygon_del_f_length}) over $len(T^{\Delta min}_c)$ or $len(T^{\Delta}_c)$, called $T^{\Delta min}_f$ or $T^\Delta_f$:
{\small
\begin{equation}
\begin{split}
len(T^{\Delta min}_f) &= \underbrace{4}_{dX_{min}} + \underbrace{5}_{dY_{min}} + \underbrace{3(N - 1)}_{dX_i} + \underbrace{3(N - 1)}_{dY_i}   \\
&= 6N + 3
\end{split}
\end{equation}}
and $6N-3$ for $T^\Delta_f$,
which is in between the upper and lower bounds on $T^{\Delta min}_c$ or $T^{\Delta}_c$. The skew in the delta values results in $T^{\Delta min}_c$ being usually better than $T^{\Delta min}_f$.

\section{Higher Base Encoding}\label{sec:higer_base_encoding}

For the next set of compression techniques we use a convenient base ($B$) transformation, $H_B(\cdot)$, to represent the encoded polygon. Encoding a large number in base 62 using alphanumeric characters [0-9A-Za-z] rather than just numeric digits [0-9] significantly reduces overall polygon string lengths. For example, one base 62 \enquote{bigit} $b$ can represent an integer up to $61$, two base-62 \enquote{bigits}, $bb$,  can represent an integer up to $3843$, while three \enquote{bigits} $bbb$ can represent an integer up to $238327$, and so on. So each delta can be represented in one or two base ($B>=62$) characters. A higher base, such as base 70, can represent even larger integers in fewer characters: a single base-70 \enquote{bigit} $b$ can represent an integer up to $69$,  two base-70 bigits $bb$ can represent an integer up to $4899$, and three base-70 bigits $bbb$ can represent  integers up to $342999$.  Likewise, $dX_{min}$ can be encoded in two base 70 bigits and $dY_{min}$ can be encoded in two or three base 70 bigits. Because of the choice of values for $X_0$ and $Y_0$, the restricted ranges and skew of $dX_{min}$  and $dY_{min}$ allow most to pack within two base-70 characters. For our analysis and experiments of compression techniques, we will use $B=70$ with all alphanumeric characters and some allowable special characters used in SMS.

String $H_{70}(T^{\Delta min}_{c})$ is the transformation of the comma-delimited values in base-70. For example, if $T^{\Delta min}_c=[9,$ $60,70]$, then $H_{70}(T_c^{\Delta min}) = $\enquote{9,y,10}. The upper bound on the length is then:
{\small
\begin{equation}
\begin{split}
len(H_{70}(T^{\Delta min}_{c})) &<= \underbrace{2}_{dX_{min}} + \underbrace{3}_{dY_{min}} + \underbrace{2(N - 1)}_{dX_i} \\
& \hspace{10mm} + \underbrace{2(N - 1)}_{dY_i} + \underbrace{2N - 1}_\text{commas} \\
&<=6N
\end{split}
\end{equation}}

and the lower bound will be:
{\small
\begin{equation}
\begin{split}
len(H_{70}(T^{\Delta min}_{c})) &>= \underbrace{1}_{dX_{min}} + \underbrace{1}_{dY_{min}} + \underbrace{(N - 1)}_{dX_i} \\
& \hspace{10mm} + \underbrace{(N - 1)}_{dY_i} + \underbrace{2N - 1}_\text{commas} \\
&>=4N-1
\end{split}
\end{equation}}

Similarly, fixed length coding $T^{\Delta min}_{f}$ in base-70 will have length strictly $4N + 1$:
{\small
\begin{equation}
\begin{split}
len(H_{70}(T^{\Delta min}_{f})) &= \underbrace{2}_{dX_{min}} + \underbrace{3}_{dY_{min}} + \underbrace{2(N - 1)}_{dX_i} + \underbrace{2(N -     1)}_{dY_i} \\
&=4N + 1
\end{split}
\end{equation}}

\section{Variable Length Encoding}\label{sec:variable}

Using the skewed distribution of the delta lengths (Figure \ref{fig:delta_diff_distribution}), we can significantly improve compression  compared to $T^{\Delta min}_{f}$, while still omitting the commas that appear in  $T^{\Delta min}_{c}$. Similar to the concept of Golomb encoding \cite{golomb}, a simple variable-length encoding of the deltas can use a single base $B$ bigit $b$ for most deltas, those below $B$, and three characters $-bb$ for the rest. For deltas greater than $B$, we use the indicator character \enquote{-} (minus) followed by two characters, $-bb$. Note that reserving \enquote{-} for this pupose means there is one less character for the base encoding.

Similarly, $dX_{min}$ and $dY_{min}$ could be encoded in one, three or four characters, using $b$, $-bb$, or $+bbb$, using $-$ or $+$ as the indicator. However,  better compression occurs if we  split $dX_{min}$ and $dY_{min}$, or equivalently $\delta X_1$, and $\delta Y_1$ for $T^\Delta$, each into two smaller parts using an agreed factor. Using the base $B$ is a particularly appropriate choice:
{\small
\begin{equation}\label{eq:dxmin}
dX_{min}=B \ast dX_{min}^\prime + dX_{min}^{\prime\prime}~\text{or}~\delta X_1=B \ast \delta X_1^\prime + \delta X_1^{\prime\prime}
\end{equation}
\begin{equation}\label{eq:dymin}
dY_{min} = B \ast dY_{min}^\prime + dY_{min}^{\prime\prime}~\text{or}~\delta Y_1 = B \ast \delta Y_1^\prime + \delta Y_1^{\prime\prime}
\end{equation}}

$dX_{min}^\prime$ and $dY_{min}^\prime$ encoding will each require at most three characters due to their distribution shown in figures \ref{fig:dxmin_distribution} and \ref{fig:dymin_distribution}. $dX_{min}^{\prime\prime}$ and $dY_{min}^{\prime\prime}$ are guaranteed to be encoded using a single bigit each.  If $B$ is 70, $dX_{min}^\prime$ will also be encoded using a single bigit. If $T^{\Delta min} = [9,60,70,73]$, then the base-70 encoding would be \enquote{9y-10-13}.

It is  important to note that we do not need commas or a fixed field to differentiate between the coordinates during decoding, since the indicator character, $-$, will suffice. The variable length encoding will be upper bounded by $6N$ for $T^{\Delta min}$ transformation, and $6N - 6$ for $T^\Delta$ transformation. While $H_{70}(T^{\Delta min}_f)$  length is better than these bounds, the skewed distributions of the NWS corpus ensure that the variable length encoding will be most often better (see Figure \ref{fig:polygon_del_f_length}).
\vspace{4mm}

\textbf{Leveraging both Skew and Limited Range of Deltas}:
We can do even better by further exploiting the skewed distributions. To improve on variable length encoding, we notice that since all deltas $dX_i, dY_i$ are less than $350$ (and $\Delta X_i, \Delta Y_i$ are less than $550$), those that are greater than $B$, and would normally use $-bb$ do not use the full range allowed by $bb$; instead we will only see $1b$, $2b$, \ldots $6b$ for $VAR^{\Delta min}$ (and $1b$, $2b$, \ldots $8b$) for $VAR^\Delta$ which allows us to replace the three character $-bb$ with a two character $xb$, where $x$ is one of several unused special characters such as $[+ * / ( ) \% \ldots ]$.  Likewise, the split $dX_{min}$ uses one char $b$ for each part. $dY_{min}^{\prime\prime}$, or $\delta Y_1^{\prime\prime}$ also uses one character. Only $dY_{min}^\prime$ or $\delta Y_1^\prime$  might use $-bb$, but their range is also restricted (no more than $167$), so instead of $-bb$, we will also use $xb$. The upper bound on the length of $T^{\Delta min}$ after applying variable length encoding ($VAR$):
{\small
\begin{equation}
\begin{split}
len(VAR_{64}^{\Delta min}) &<= \underbrace{2}_{dX_{min}^\prime~\&~dX_{min}^{\prime\prime}} +
\underbrace{3}_{dY_{min}^\prime~\&~dY_{min}^{\prime\prime}}  \\
&  \hspace{10mm} + \underbrace {2(N - 1)}_{dX_{i}} + \underbrace{2(N - 1)}_{dY_{i}}  \\
&<= 4N + 1
\end{split}
\end{equation}}

Note again that the base $64$ reserves an additional six special characters for the allowed $70$. Transformation $T^\Delta$ needs two more special characters due to the bounds on $(\Delta X_i, \Delta Y_i)$, $len(VAR_{62}^{\Delta})$ is upper bounded by $4N-3$. The lower bound on the length of $VAR^{\Delta min}_{64}$ is achieved when all values are less than $B=64$ (and less than $B=62$ for $VAR^{\Delta}_{62}$)\footnote{One could choose to allocate less than 6 or 8 characters to the $xb$, but then the "-" would be needed in a  few cases}:
{\small
\begin{equation}
\begin{split}
len(VAR_{64}^{\Delta min}) &>= \underbrace{2}_{dX_{min}^\prime~\&~dX_{min}^{\prime\prime}} + \underbrace{2}_{dY_{min}^\prime~\&~dY_{min}^{\prime\prime}} \\ 
& \hspace{10mm} + \underbrace{(N - 1)}_{dX_{i}} + \underbrace{(N - 1)}_{dY_{i}}  \\
&>= 2N + 2
\end{split}
\end{equation}}
For example, $VAR_{64}^{\Delta min} = $\enquote{9y+11}, if: \\ $T^{\Delta min} = [9,60,62,65]$. The lower bound for $VAR_{62}^\Delta$ remains same as for $VAR_{64}^{\Delta min}$.

\section{Bignum Compression}\label{sec:big}
\textit{Bignum} compression further improves on $T^{\Delta min}$ or $T^{\Delta}$, by combine each delta pair $(dX_i, dY_i)$ or $(\Delta X_i, \Delta Y_i)$ into a larger single number:
{\small
\begin{equation}\label{eq:pair_small}
dXY_i = dX_i \ast XX + dY_i,
\end{equation}}
where $XX$ can be a fixed choice or chosen based on the range of $dY_i$ to make sure there is \enquote{space} for $dY_i$. For instance based on our corpus $XX = 350$ will be an appropriate value to \enquote{make space} for $dY_i$ because $dY_i$ is less than $350$ for the NWS corpus; similarly $\Delta Y_i$ is always less than $550$.

Likewise, we can combine $(dX_{min}, dY_{min})$ or $(\delta X_1, \delta Y_1)$ using a larger factor:

{\small\begin{equation}\label{eq:pair_big}
dXY_{min} = dX_{min} \ast Y_{factor} + dY_{min}~,
\end{equation}}
where $Y_{factor}$ can be chosen based on the range of $dY_{min}$. An appropriate value based on the corpus is $10,000$ to make space for the $dY_{min}$.

Expanding on this pair of deltas idea, we can aggregate all deltas into a single large integer, using a simplified form of arbitrary precision integer arithmetic. The large integer is computed by successive pairing of elements of $T^{\Delta min}$:
{\small
\begin{equation}\label{eq:big_del}
BIG_{i+1}^{\Delta min} = (BIG_i^{\Delta min} \ast XX^2) + ((dX_i + 1)\ast XX) + (dY_i + 1),
\end{equation}}
where $i \in [1, N-1]$, and $BIG_i^{\Delta min}$ is 0. We add one to all $dX_i$ and $dY_i$ values to avoid the pathological case when delta values are zero.\footnote{Strictly speaking, only the first value, $dX_1$ could cause a problem.} For $T^\Delta$ we use essentially the same equation: 
{\small
\begin{equation}
BIG_{i+1}^{\Delta} = (BIG_i^{\Delta} \ast XX^2) + ((\Delta X_i + 1)\ast XX) + (\Delta Y_i + 1),
\end{equation}}
where $i \in [1, N-2]$. The value 
$XX$ is chosen for each polygon using a "indicator" character $S$ defined by:
{\small
\begin{equation}\label{eq:big_s}
S=\frac{max(\sup_i dX_i,\sup_i dY_i)}{d} + 1
\end{equation}}

{\small
    \begin{equation}
XX = d\ast S + 1
\end{equation}}

The value for $d$ can be chosen based on the distribution of the deltas. For the NWS corpus, $d=6$ ensures $XX$ to be large enough to encode any $(dX_i, dY_i)$, while  $d=9$ ensures $XX$ will be large enough for $(\Delta X_i, \Delta Y_i)$.  Also, the above selection of $XX$ guarantees its encoding via $S$ using a single character in base-70 for $S$. \footnote{A slightly better result is obtained if we  use a piecewise linear approximation of $XX(S)$ to $max(\sup_i dX_i,\sup_i dY_i)$, whereby we exactly match for $XX=0:30$, and then more granular from $XX=31:330$ (or $31:540$ for $T^\Delta$).}

Larger values for the starting points in $T^{\Delta min}$ and $T^{\Delta}$ could also be included in $BIG$ in several ways. Firstly by choosing an appropriate set of factors based on the distribution, such as $X_{factor}=3500$ and $Y_{factor}=10000$ for NWS corpus, and then apply the following:  
{\small \begin{equation}\label{eq:big_dmin}
BIG_{N}^{\Delta min} = (BIG^{\Delta min}_{N-1}\ast X_{factor} + dX_{min})\ast Y_{factor} + dY_{min}
\end{equation}
\begin{equation}\label{eq:big_diff}
BIG_{N-1}^{\Delta} = (BIG^{\Delta}_{N-2}\ast X_{factor} + \delta X_{1})\ast Y_{factor} + \delta Y_{1}
\end{equation}}
The $X_{factor}$ and $Y_{factor}$ make space for their $(dX_{min},$ $dY_{min})$ or $(\delta X_1, \delta Y_1)$. The encoded string in base-70 representation is a concatenation: 
                 \[\overline{BIG}_{70}^{\Delta min} = S~\bullet~BIG_{N}^{\Delta min}\]
                 \[\overline{BIG}_{70}^{\Delta} = S~\bullet~BIG_{N-1}^{\Delta}\]

An estimate of the bounds on $BIG$ can be obtained by noticing that we are essentially placing each $dX_i$ and $dY_i$, or $\Delta X_i$ and $\Delta Y_i$ into an $XX$ sized space, essentially \enquote{shifting} by $log_2(XX)$ bits, concatenating into a big number, and then chopping into $B$ sized characters (each $log_2(B)$ bits). Thus allowing one character for $S$ and about four characters for the $dX_{min}$, $\delta X_1$, etc. shifted by $X_{factor}$ and $Y_{factor}$, we get approximately: 
{\small
\begin{equation}
\begin{split}
    len(\overline{BIG}^{\Delta min}_B) & \approx 1 + \frac{(log_2(X_{factor}) +log_2(Y_{factor})}{log_2(B)} \\
& \hspace{10mm}+ \frac{2(N-1)log_2(XX)}{log_2(B)}\\
len(\overline{BIG}^{\Delta min}_{70})& \approx 2(N-1)\frac{log_2(XX)}{log_2(70)} + 5.09
\end{split}
\end{equation}}
\comm{AJ: To validate constants}
and
{\small
\begin{equation}
\begin{split}
len(\overline{BIG}^{\Delta}_B) & \approx 1 + \frac{(log_2(X_{factor}) +log_2(Y_{factor})}{log_2(B)}\\
& \hspace{10mm} + \frac{2(N-2)log_2(XX)}{log_2(B)}\\
len(\overline{BIG}^{\Delta}_{70})& \approx 2(N-2)\frac{log_2(XX)}{log_2(70)} + 5.09
\end{split}
\end{equation}}

for $B=70$, $X_{factor}=3500$ and $Y_{factor}=10000$.

Thus when $XX=B$,  this is essentially one more than the bound on $VAR$, and gets increasingly better as $XX$ decreases below $B$. Furthermore, at larger $XX$, $BIG$ will be better than $VAR$ for certain polygons when a significant number of the delta coordinates in the polygon are larger than $B$, thus requiring the longer two-character $xb$ representation.
Note that actual encoding base B is smaller for $VAR$ than for $BIG$ for the same set of available characters.

The best case compression in the $BIG$ technique for a polygon would be to pick the compression parameters adaptively as follows:
\[XX=max(\sup_i dX_i,\sup_i dY_i)+2\footnote{Since we add one to all deltas in Eq.\ref{eq:big_del}, $XX$ has to be strictly greater than all $dX_i + 1$ and $dY_i + 1$ such that we can decode correctly. Strictly, we only have to deal with the first point specially.}\]
\[X_{factor} = dX_{min}+1\]
\[Y_{factor} = dY_{min}+1\]

However, outputting these exact choices as part of the string would add too many characters. Indeed, we could then directly encode $dX_{min}$ and $dY_{min}$ in other ways but experiments suggest that will not be any better than encoding using the $X_{factor}$ and $Y_{factor}$ approach. \footnote{We can also apply the same $VAR$ splitting approach, treating the parts of $dX_{min}$ and $dY_{min}$   as four additional deltas. Thereafter, using a potentially larger $XX^\prime$ in place of $XX$ in Eq.\ref{eq:big_del}, and also in Eq.\ref{eq:big_dmin} in place of $X_{factor}$ and $Y_{factor}$. 
This adaptively picking parameters holds true for $(X_1^\prime, Y_1^\prime)$. However, in the NWS corpus it is not as good as the $X_{factor}$, $Y_{factor}$ approach.}

\begin{table*}
\begin{tabular*}{\textwidth}{ll@{\extracolsep{\fill}}l}
\toprule
\small{Transformation} & \small{Variable} & \small{Value} \\
\toprule
$T^{\Delta min}$ & $O$ & [31.3,-97.4,31.51,-97.55,31.8,-96.99,31.58,-96.84,31.3,-97.4] \\
&$O^\prime$    & [3130,9740,3151,9755,3180,9699,3158,9684]  \\\cmidrule{2-3}% That's the rule you're looking for.
&$T^{\Delta min}$    & [1530,3684,0,56,21,71,50,15,28,0] \\
    &$T^{\Delta min}_c$  & \enquote{1530,3684,0,56,21,71,50,15,28,0} \\
    &$T^{\Delta min}_f$  & \enquote{153003684000056021071050015028000} \\
    &$\overline{BIG}^{\Delta min}$ & \enquote{14818307150871153
    03684} \\
    &$\overline{BIG}_{70}^{\Delta min}$ & \enquote{Z7YfAH*`vmYi4} \\
\midrule
$T^{\Delta}$ & $O$ & [31.3,-97.4,31.51,-97.55,31.8,-96.99,31.58,-96.84,31.3,-97.4] \\
&$O^\prime$    & [3130,9740,3151,9755,3180,9699,3158,9684]  \\\cmidrule{2-3}% That's the rule you're looking for.
&$T^{\Delta}$    & [1530,3740,42,30,58,111,43,29] \\
    &$T^{\Delta}_c$  & \enquote{1530,3740,42,30,58,111,43,29} \\
    &$T^{\Delta}_f$  & \enquote{153003740042030058111043029} \\
    &$\overline{BIG}^{\Delta}$ & \enquote{33202964332840303740} \\
    &$\overline{BIG}_{70}^{\Delta} $ & \enquote{ZBqu20DM8m*y} \\
\bottomrule
\end{tabular*}
\caption{Example of compression with different transformations. Notice that $T^\Delta$ is always less in length in comparison to $T^{\Delta min}$, primarily because it has one less point. Base 70 is considered for convenience and easy comparison with $VAR$.}
\end{table*}

\section{Variable Length Encoding\\ with Repeated Substring Dictionary (RSD)}\label{rsd}
Here we extend the idea of variable length encoding further by usage of a dictionary\footnote{A dictionary contains a set of mapping objects (key, value)}. The input string to this technique is the transformed list of coordinates represented by $VAR_{B}^{\Delta min}$ or $VAR_{B}^{\Delta}$. As indicated above, each delta will either be a single $b$ or two character $xb$.

Inspired by LZW, we exploit the statistical redundancy in polygons across the corpus to  generate a static dictionary for the entire NWS corpus, and provide that to the encoding and decoding systems. This dictionary is essentially a set of most frequently repeated three character sub-strings. We are using the same base $B$ as for variable length encoding ($VAR$) in the dictionary for keys and values\footnote{Actually, we can use the full allocated 70 character size for the table}. A sub-string of size two would not have performed any extra compression since we need to prefix a dictionary value with an indicator character to make the distinction of dictionary value in the encoding, and non-dictionary value. The size of the dictionary is constrained by the available set of characters, and any encoding using the dictionary will be of length two. We tried two approaches to construct the dictionary:

\begin{itemize}
\item Fixed field matching: Chop the character string into disjoint three character substrings, and keep a count of each unique sub-string.
\item Sliding window: Slide a three character window across the string and keep a count of each substring in the dictionary. Again, the size is restricted by the base. 
\end{itemize}

The fixed field case is easier to implement, and faster to execute, but the sliding window gives better results.

There may be cases when none of the repeated substrings occur in the input, and no extra compression is achieved, but there is no penalty, other than a linear order of cost of creating and storing the dictionary, and the finding any matching substrings. For any variable length encoding using base $B$, $VAR_B$, the encoding with RSD will be referred as $VAR_{B-1\_RSD}$. $B-1$ due to the extra special character for encoding substrings found in the dictionary.

Table \ref{tab:rsd} is an example of a static dictionary storing 70 three character substrings for $VAR_{63}^{\Delta min}$ encodings of NWS corpus. $000$ occurred $1414$ times, whereas $00N$ was the last entry in the table which occurred $96$ times:

\vspace{4mm}
\begin{table}[h!]
\centering
\begin{tabular}{||c c||} 
\hline
key & base-70 value \\ [0.5ex] 
\hline\hline
000 & 0 \\
100 & 1 \\
... & ... \\
C00 & N \\
... & ... \\
00N & \_ \\ [1ex] 
\hline
\end{tabular}
\caption {Repeated Substring Dictionary for variable length encoding} \label{tab:rsd} 
\end{table}

\begin{table*}
\begin{tabular*}{\textwidth}{ll @{\extracolsep{\fill}} l}
\toprule
\small{Transformation} & \small{Variable} & \small{Value} \\
\toprule
$T^{\Delta min}$ & $O$ & [30.97,-92.28 30.89,-92.04 30.61,-92.22 30.65,-92.34 30.97,-92.28] \\
&$O^\prime$    & [3197,9228,3089,9204,3061,9222,3065,9234]  \\\cmidrule{2-3}% That's the rule you're looking for.
&$T^{\Delta min}$    & [1461,3204,36,24,28,0,0,18,4,30] \\
    &$VAR_{64}^{\Delta min}$ & \enquote{Mro4aOS00I4U} \\
    &$VAR_{63\_RSD}^{\Delta min}$ & \enquote{NCosaOS@v4U} \\
\midrule
$T^{\Delta}$ & $O$ & [30.97,-92.28 30.89,-92.04 30.61,-92.22 30.65,-92.34 30.97,-92.28] \\
&$O^\prime$    & [3197,9228,3089,9204,3061,9222,3065,9234]  \\\cmidrule{2-3}% That's the rule you're looking for.
&$T^{\Delta}$    & [1497,3228,15,47,55,36,8,24] \\
    &$VAR_{62}^{\Delta}$ & \enquote{O9q4Flta8O} \\
    &$VAR_{61\_RSD}^\Delta$ & \enquote{OXquFlta8O} \\

\bottomrule
\end{tabular*}
\caption{Example of compression with different transformations. Notice that @ in $VAR_{63\_RSD}^{\Delta min}$ is the indicator character to distinguish between a dictionary value and non-dictionary value. Although, $VAR_{61}^{\Delta}$ had no keys in its RSD dictionary, and therefore $VAR_{61\_RSD}^\Delta \approx VAR_{62}^\Delta$}
\end{table*}

\section{Arithmetic Encoding}\label{ae}
Arithmetic encoding (AE) is a variable length and lossless encoding technique. For compression and decompression AE relies on a probabilistic model. The algorithm is recursive for each character i.e. it operates upon and encodes (decodes) one data symbol per iteration \cite{langdon1984introduction}.

The probability model over the possible characters to perform the encoding and decoding steps is essential for optimal compression. Specifically, the compression ratio depends on how well the probability model represents the string of characters to be encoded. For our experiments with polygons the probability of occurrence of any character is based on the entire corpus of polygons. 

For the purpose of the polygons, we will define the character sequence $\Sigma  = (0 1 2 3 4 5 6 7 8 9 )$.
Before applying AE, all polygons were transformed to deltas by the same heuristics used to get the $T^{\Delta min}$ string.

Arithmetic encoding is applied to this delta string. The basic algorithm is described below.
\vspace{4mm}

\begin{enumerate}[1:]
\item Begin with the current interval $[l_0, h_0)$ initialized to $[0, 1)$.
\item Sub divide the current interval $[l_0, h_0)$ proportional to the probability of each character in $\Sigma$.
\item For each character $c_i$ of the polygon string, we perform two steps:
Consider the probability interval for $c_i$, call it $[l_i, u_i)$ and make it the current interval.
Subdivide the current interval into subintervals, one for each possible character, and defined by probabilities over $\Sigma$.
\item We output enough bits representing the final interval $[l_n, u_n)$, where $n$ is the length of the polygon string.
\end{enumerate}

The output from step 3 of the algorithm is a binary representation of any real value in the interval $[l_n, u_n)$. A real value in the final interval uniquely identifies a string of characters provided the length of the string to be decoded is known by the decoder. In other words, each input string generates a unique probability interval due to the recursive approach of dividing the probability intervals.

As stated before, we need to embed the length of the input string for the decoder to retrieve the original polygon but here we embed the number of coordinates of the polygon in the compressed string which is sufficient for decompression.

\section{Standard Methods-LZW,\\ Golomb, Huffman, 7zip, gzip} \label{sec:others}
LZ78 \cite{ziv1977universal} is a variant of the LZW (Lempel-Ziv-Welch) algorithm, implemented for example in the well-known GZIP. The basic idea of the LZW algorithm is to take advantage of repetition of substrings in the data \cite{blackstock}  and use a smaller length encoding for such repetitions using data structure like a dictionary with one-to-one mapping of substrings to encodings. We tried the LZW algorithm with both $T_c^{\Delta min}$ and $T_c^{\Delta}$ strings.

With $T^\Delta$ as an input, we also tried other standard string compression algorithms available \cite{matt} like \textit{7zip}, and \textit{gzip}, but their compressed lengths were not as good as our $BIG$ or $VAR$ encoding techniques.

We compared our results to Golomb coding\cite{golomb}, which is well studied technique for input values following a geometric distribution, essentially where most values are small, like our delta distributions. Golomb is essentially a concatenation of a variable length unary coded prefix (a string of $1$ followed by a $0$) and a fixed bit length remainder. We then encoded this bit string in base $B$ characters for both transformations.

We also compared our results to Huffman encoding \cite{huffman1952method} using probabilities similar to the $AE$ method. Huffman builds a coding tree using these probabilities, but due to its size ($4835$ leaf nodes for $\Delta min$ and $4715$ leaf nodes for $\Delta$ transformation) this compression technique will not be space efficient on a mobile phone.

\section{Practical Considerations}\label{sec:practical}
In order to embed a compressed polygon string in a WEA message, or a simulated WEA message for the trials, we need to signal the start and stop of the polygon string, or the start and length, as appropriate. In most cases, we could prefix, \# or \#p and a postfix, \# or \#]. In order to embed the polygon in longer text messages that might contain a \# character, we encode any other \# as a \#\#. Furthermore, as indicated, we use base 70 (or higher) to compress numeric strings. We can use a  larger base, such as 90, however we need to limit to only use characters that can be included in an SMS or broadcast message, and exclude any characters that are also used by the compression scheme (such as the \# sentinel, the -,  + and other characters used for signs for variable length $VAR$ substrings,  and the @ indicator in the RSD approach).  Thus in most cases, the embedded polygon will be 2-4 characters longer than the numbers indicated above. 

\begin{figure*}[!ht]
  \centering
  \includegraphics[width=120mm]{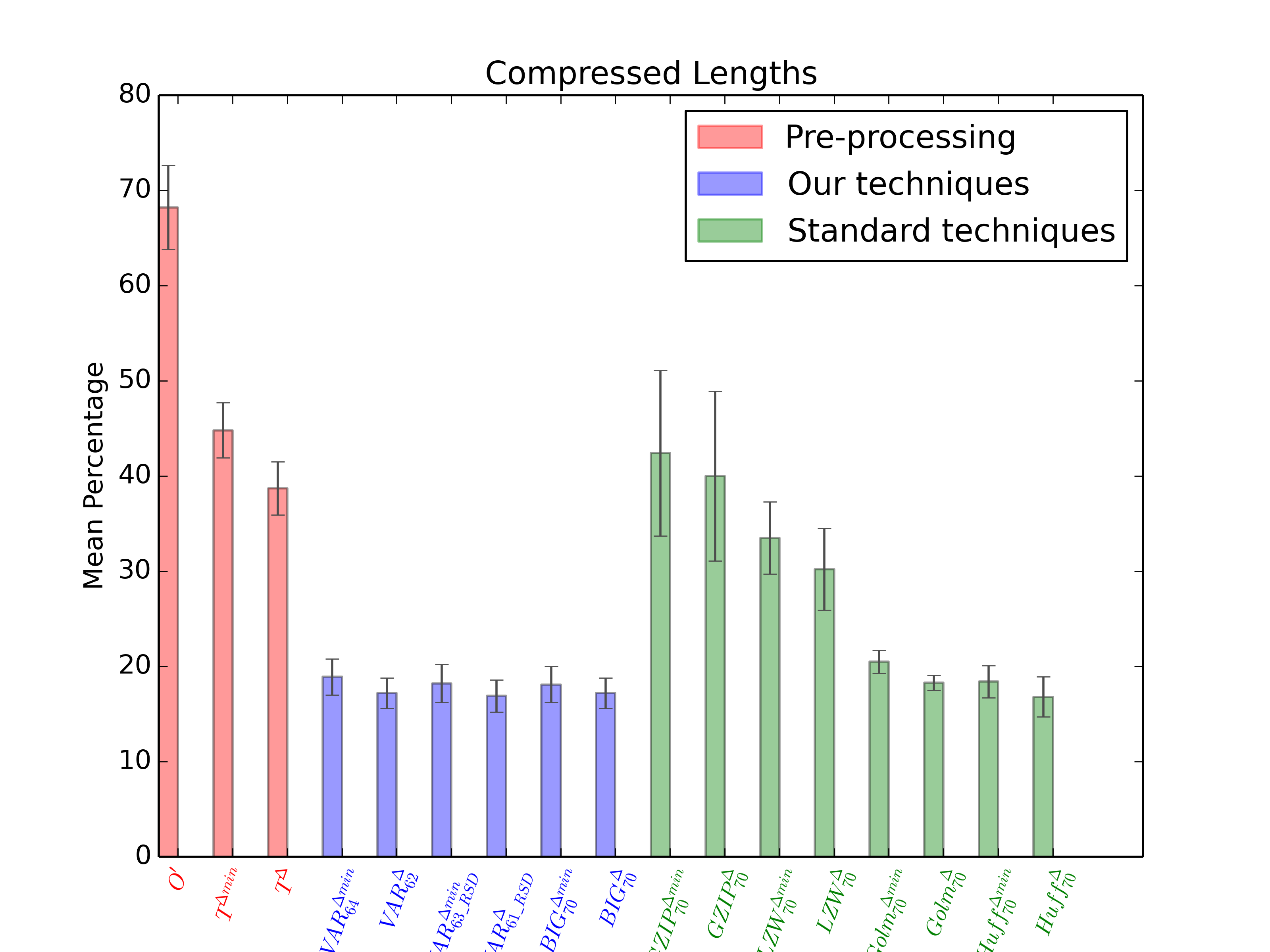}
    \caption{Summary of Base 70 results showing mean percentage of compressed length. Our compression techniques (shown in blue) have mean compressed length less than $20$ with low error (standard deviation) in comparison to some standard techniques (shown in green).}
    \label{fig:base_70_results}
\end{figure*}

Also for Arithmetic Encoding (AE) although we need to include the number of characters to be retrieved by the decoder, we will need a prefix and a postfix as sentinels. For some values of $S$ in BIG, we can save one character by using a different sentinel, \#q, \#r, \#s instead of \#pS.

See \cite{etsigsm} for discussion of character sets.  Sometimes, operator gateways used in our experiments would not transmit some characters, and so we used base 70, even though a higher base would somewhat improve the compression percentages.

\section{Polyalgorithm}
Because of the variability in the results for each technique, each has some polygons for which it is the best method, leading to consideration of combinations of techniques  and adaptive technique selection.

As indicated below, the two best techniques $BIG^{\Delta}(S)$ and $VAR^{\Delta}_{RSD}$ are close in output character lengths. We can create a combined technique,  $POLY^{\Delta}(S)$, that uses $BIG^{\Delta}(S)$ where it is best and $VAR^{\Delta}_{RSD}$ where it is best, adding an extra character $S=0$ to signal $VAR^{\Delta}_{RSD}$ and other values of $S$ to signal and control $BIG^{\Delta}(S)$. Since $BIG^{\Delta}$  and $VAR^{\Delta}_{RSD}$ are so close (with $BIG$ better for small $XX$),  adding this extra character can swamp the benefit. This extra character can be saved in the practical case by using a different sentinel \#q, instead of \#p0. Note that if we decide that only 70 characters are available, we use the full $B=70$ as encoding base for $BIG$, but since we are reserving 9 characters for indicators in  $VAR^\Delta_{RSD}$, we actually use only $B=61$ as encoding base for the corresponding $VAR^\Delta_{61,RSD}$. Thus $VAR$, and $VAR_{RSD}$ do \enquote{waste} some of the available characters.

\comm{Some graph or chart to illustrate this}

It is important to note that $BIG$ should always be better than $VAR$ when $dXY_{max}=\sup_i(dX_i,dY_i)  < B$\footnote{This is the B used in the comparible VAR encoding}, which, as can be seen from Figure \ref{fig:delta_diff_distribution}, occurs more than 65\% of the time for $B=62$ and more than 80\% for $B=90$ for $T^{\Delta min}$, and when $\Delta XY_{max}=\sup_i(\Delta X_i,\Delta dY_i) < B$ which occurs 34\% for $B=62$ and 62\% for $B=90$ for $T^{\Delta min}$.  Typically the shorter length of $BIG$ occurs when $XX$ is quite a bit less than $B$.

So in setting $XX(S)$ effectively for the polyalgorithm, its important to have $XX$ close to the $dXY_{max}$ or $\Delta XY_{max}$. In many cases $BIG$ will be better than $VAR$, and will be substantially better for smaller $XX$ and for those polygons when many of their deltas require two base $B$ characters for larger $XX$.

\begin{figure*}
\captionsetup{justification=centering}
\begin{subfigure}{.45\textwidth}
  \centering
  \includegraphics[width=\linewidth]{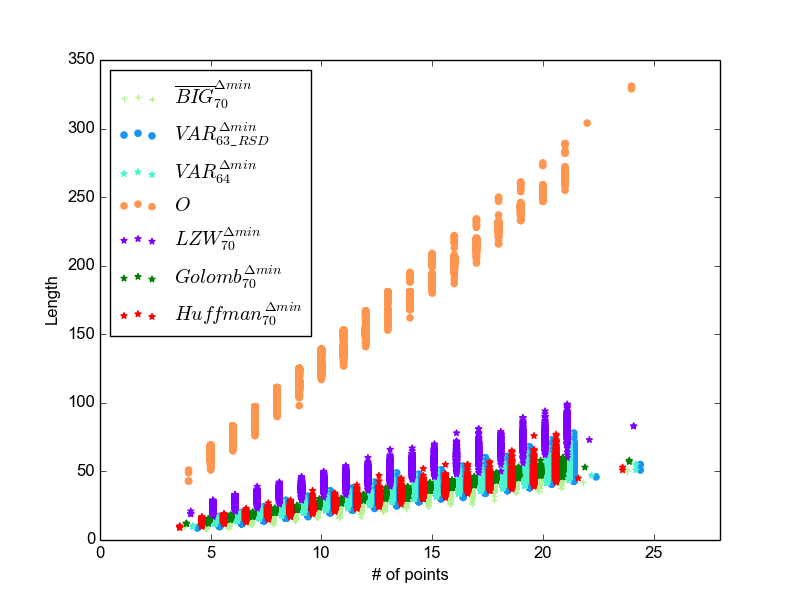}
    \caption{Compression lengths with $T^{\Delta min}$ as input transformation}
    \label{fig:all_del}
\end{subfigure}
\begin{subfigure}{.45\textwidth}
  \centering
    \includegraphics[width=\linewidth]{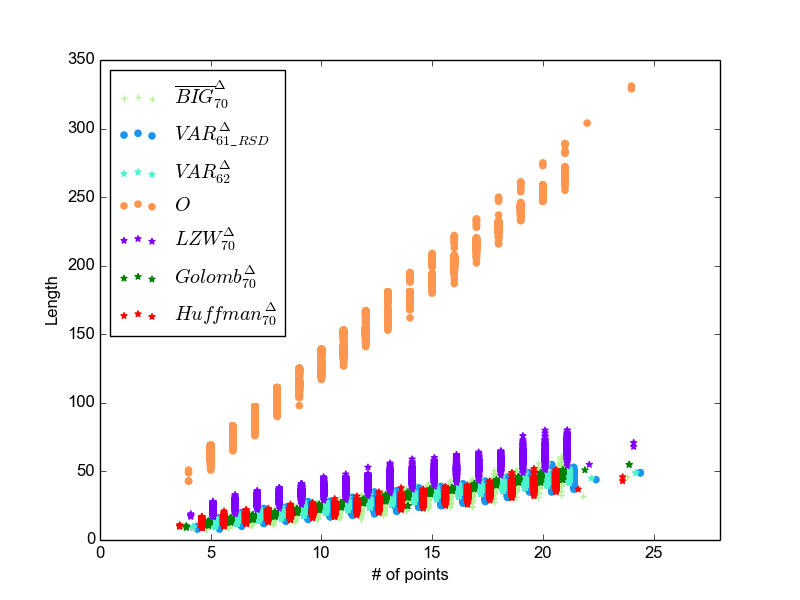}
    \caption{Compression lengths with $T^{\Delta}$ as input transformation}
    \label{fig:all_diff}
\end{subfigure}
\caption{Comparison of compressed polygon lengths with original lengths $(O)$ of polygon strings}
\end{figure*}

\section{Comparison, Discussion \& Conclusions}
Figure \ref{fig:base_70_results} summarizes the 70 character (usually base 70, except for the $VAR$ techniques) compression results of the 11370 polygons collected from the NWS online portal for 2012, 2013, and through December, 2014. Base 70 was chosen to incorporate the reserved characters used by $VAR$ and $VAR_{RSD}$ apart from the alphanumeric characters of base 62. We noticed small improvements at a higher value 90, particularly in the maximum values of lengths, and the reduced standard deviation.

In general, $VAR$ can be characterized as \emph{the best} and simplest direct technique, with the best compression ratio and  least variability.  $BIG$ is very close, and as indicated above is better when $XX<B$ and when a significant number of deltas require two characters in $VAR$ for larger $XX$.  However $VAR_{RSD}$ is slightly better overall. Because the results are so close, which method is ultimately deemed best depends strongly on the specific polygon and the overall distribution of the polygons.

Figures~\ref{fig:all_del},~\ref{fig:all_diff} shows that all of the methods yield substantial compressions. As seen from the figures and from the formulas displayed earlier, the results are essentially linear in $N$. 

$BIG^\Delta$ and $VAR^\Delta$ are the best direct techniques, each leading in about 50\% of the cases. We observed from the detailed results for each polygon that $VAR^\Delta_{61\_RSD}$ is better than $VAR^\Delta_{62}$, and for more than 50\% of the time $VAR^\Delta_{61\_RSD}$ is better than $BIG_{70}$. Thus we introduce the polyalgorithm, $POLY_{70}$ which is slightly better than  $VAR^\Delta_{61\_RSD}$ overall.

\begin{figure*}
\captionsetup{justification=centering}
\begin{subfigure}{.45\textwidth}
    \centering
    \includegraphics[width=\linewidth]{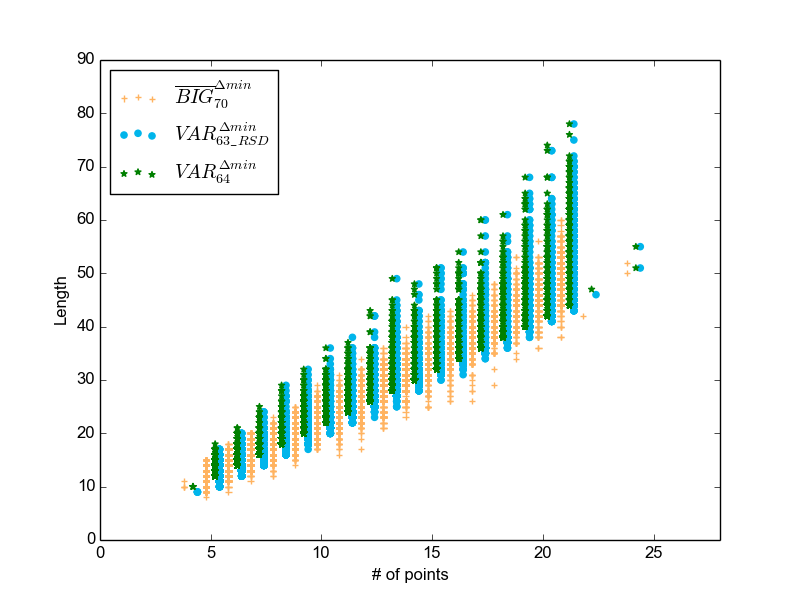}
    \caption{Compression lengths with $T^{\Delta min}$ as input transformation}
    \label{fig:del_results}
\end{subfigure}
\begin{subfigure}{.45\textwidth}
    \centering
    \includegraphics[width=\linewidth]{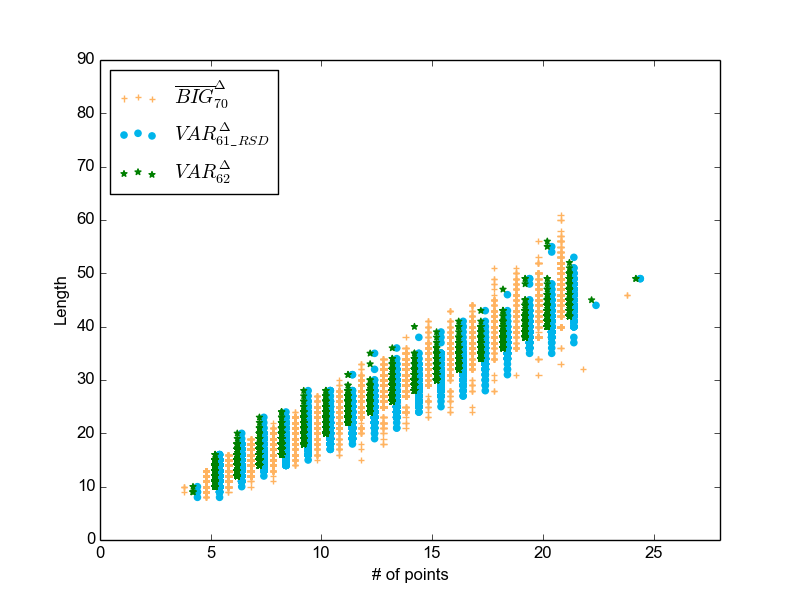}
    \caption{Compression lengths with $T^{\Delta}$ as input transformation}
    \label{fig:diff_results}
\end{subfigure}
\caption{Comparison of best techniques}
\end{figure*}

Figures \ref{fig:del_results} and \ref{fig:diff_results} compare the best methods using a set of 70 available characters, with corresponding encoding base $B=61,62,63,64$, or $70$ for the different techniques, but we would get similar results with a larger base. 

\begin{figure*}
\captionsetup{justification=centering}
\begin{subfigure}{.45\textwidth}
  \centering
  \includegraphics[width=\linewidth]{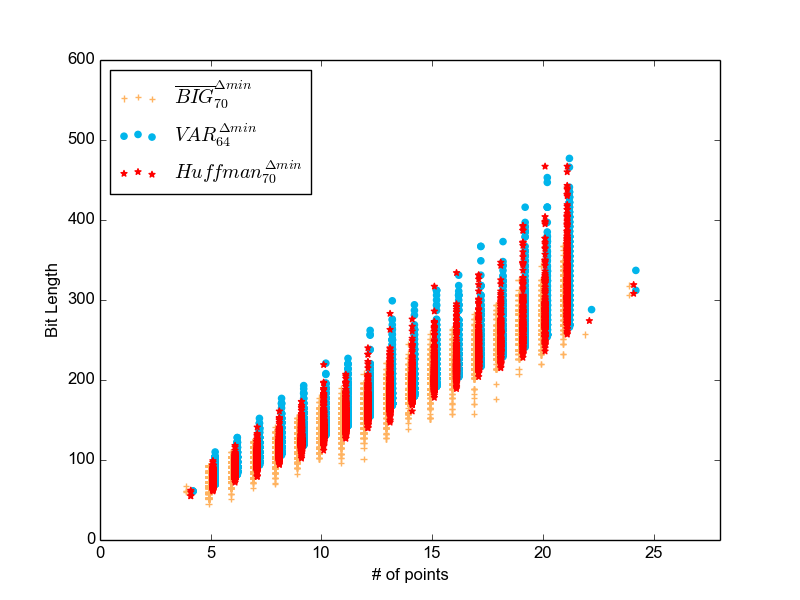}
    \caption{Compression lengths in bits with $T^{\Delta min}$ as input transformation}
    %\label{fig:polygon_del_c_length}
\end{subfigure}
\begin{subfigure}{.45\textwidth}
  \centering
    \includegraphics[width=\linewidth]{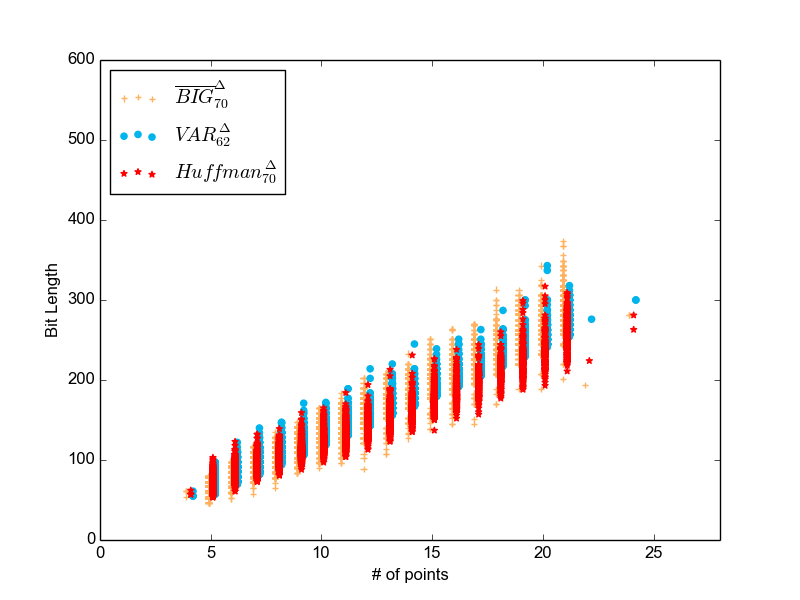}
    \caption{Compression lengths in bits with $T^{\Delta}$ as input transformation}
    %\label{fig:polygon_del_f_length}
\end{subfigure}
\caption{Comparison of compressed polygon lengths with Shannon bound}
\label{fig:huffman_results}
\end{figure*}

Compression of polygons using kd-trees \cite{devillers2000geometric} is known to have good compression ratios only when the polygon mesh is sparse with few edges \cite{alliez2005recent}. 

As we explored various techniques, we experimented with several small optimizations that would occasionally save a character. These involved adjusting some of the parameters such as $X_0, Y_0, X_{factor},$ and $Y_{factor}$, changing the piece-wise linear representation of $XX(S)$ and so forth, but overall the parameters presented in this paper seemed the best compromise.

By Shannon's coding theorem \cite{shannon2001mathematical} optimal compression for a set $\Sigma$ of symbols in which each symbol $c$ has the probability of occurrence $p_c$, then the entropy is given by:
{\small
\begin{equation}
E(\Sigma) = \sum_{c \in \Sigma} -p_c \cdot \lg_2(p_c)
\end{equation}}
bits from encoding each symbol. Huffman encoding is an optimal prefix encoding technique, such that the Huffman encoding length is never longer than the entropy of the given distribution. In figure \ref{fig:huffman_results}, we see that $BIG$ and $VAR$ are close to the almost optimal Huffman encoding. None of the other techniques like \textit{7zip}, LZW, and \textit{gzip} were as good on the NWS corpus as these two. 

\section{Future Work}
Future work includes exploring even more use of statistical skew, such as an extended form of variable length encoding. This approach will allow us to take the variable-length encoding strategy further by starting from a base-2 (binary) representation, and using one or two bits to determine the length of the following field. We will further explore this more Golomb-like option.

We also plan to use integer programing to find the near optimal set of compression parameters such that we maximize the leverage of any skew in the delta values.
\newpage
\section{Acknowledgements}
We gratefully appreciate support from the Department of Homeland Security, Science and Technology Directorate. We also appreciate the conversations, advice and weather polygon data supplied by Mike Gerber of the National Oceanic and Atmospheric Administration's National Weather Service.

\bibliographystyle{abbrv}
\bibliography{main}

\end{document}